\documentclass[11pt]{article}
\usepackage{amsmath,amssymb,amsthm,amsxtra,overpic,bbm,bm,epsfig}
\def\thefootnote{\fnsymbol{footnote}}

\begin{document}

\vspace{0.2cm}

\begin{center}
{\Large\bf Towards testing the unitarity of the $3\times 3$ lepton
flavor mixing matrix in a precision reactor antineutrino oscillation
experiment}
\end{center}

\vspace{0.1cm}

\begin{center}
{\bf Zhi-zhong Xing}
\footnote{E-mail: xingzz@ihep.ac.cn} \\
{Institute of High Energy Physics, Chinese Academy of
Sciences, P.O. Box 918, Beijing 100049, China, \\
Center for High Energy Physics, Peking University, Beijing 100080, China, \\
Theoretical Physics Center for Science Facilities, Chinese Academy
of Sciences, Beijing 100049, China}
\end{center}

\vspace{1.5cm}

\begin{abstract}
The $3\times 3$ Maki-Nakagawa-Sakata-Pontecorvo (MNSP) lepton flavor
mixing matrix may be slightly non-unitary if the three active
neutrinos are coupled with sterile neutrinos. We show that it is in
principle possible to test whether the relation $|V^{}_{e1}|^2 +
|V^{}_{e2}|^2 + |V^{}_{e3}|^2 =1$ holds or not in a precision
reactor antineutrino oscillation experiment, such as the recently
proposed Daya Bay II experiment. We explore three categories of
non-unitary effects on the $3\times 3$ MNSP matrix: 1) the {\it
indirect} effect in the (3+{\bf 3}) flavor mixing scenario where the
three heavy sterile neutrinos do not take part in neutrino
oscillations; 2) the {\it direct} effect in the (3+$\mathbbm{1}$)
scenario where the light sterile neutrino can oscillate into the
active ones; and 3) the {\it interplay} of both of them in the
(3+$\mathbbm{1}$+{\bf 2}) scenario. We find that both the
zero-distance effect and flavor mixing factors of different
oscillation modes can be used to determine or constrain the sum of
$|V^{}_{e1}|^2$, $|V^{}_{e2}|^2$ and $|V^{}_{e3}|^2$ and its
possible deviation from one, and the active neutrino mixing angles
$\theta^{}_{12}$ and $\theta^{}_{13}$ can be cleanly extracted even
in the presence of light or heavy sterile neutrinos. Some useful
analytical results are obtained for each of the three scenarios.
\end{abstract}

\begin{flushleft}
\hspace{0.8cm} PACS number(s): 14.60.Pq, 13.10.+q, 25.30.Pt \\
\hspace{0.8cm} Keywords: non-unitary effects,
sterile neutrinos, reactor antineutrino oscillations
\end{flushleft}

\def\thefootnote{\arabic{footnote}}
\setcounter{footnote}{0}

\newpage

\section{Introduction}

The $\nu^{}_e$, $\nu^{}_\mu$ and $\nu^{}_\tau$ neutrinos are {\it
active} in the sense that they take part in the standard weak
interactions. They are significantly different from their
corresponding mass eigenstates $\nu^{}_1$, $\nu^{}_2$ and $\nu^{}_3$
as a result of non-degenerate neutrino masses and large lepton
flavor mixing \cite{PDG}. Whether there exist the {\it sterile}
neutrinos, which do not directly take part in the standard weak
interactions, has been an open question in particle physics and
cosmology. One is motivated to consider such ``exotic" particles for
several reasons. On the theoretical side, the canonical (type-I)
seesaw mechanism \cite{SS} provides an elegant interpretation of the
small masses of $\nu^{}_i$ (for $i=1,2,3$) with the help of two or
three heavy sterile neutrinos, and the latter can even help account
for the observed matter-antimatter asymmetry of the Universe via the
leptogenesis mechanism \cite{FY}. On the experimental side, the LSND
\cite{LSND}, MiniBooNE \cite{M} and reactor \cite{R} antineutrino
anomalies can all be explained as the active-sterile antineutrino
oscillations in the assumption of one or two species of sterile
antineutrinos whose masses are below 1 eV \cite{Schwetz}.
Furthermore, a careful analysis of the existing data on the Big Bang
nucleosynthesis \cite{Mangano} or the cosmic microwave background
anisotropy, galaxy clustering and supernovae Ia \cite{Raffelt} seems
to favor at least one species of sterile neutrinos at the sub-eV
mass scale. On the other hand, sufficiently long-lived sterile
neutrinos in the keV mass range might serve for a good candidate for
warm dark matter if they were present in the early Universe
\cite{Bode}. That is why the study of sterile neutrinos becomes a
popular direction in today's neutrino physics \cite{Review}.

In the presence of small active-sterile neutrino mixing, the
conventional $3\times 3$ Maki-Nakagawa-Sakata-Pontecorvo (MNSP)
lepton flavor mixing matrix \cite{MNS} is just the submatrix of a
$(3+n) \times (3+n)$ unitary matrix $V$ which describes the overall
flavor mixing of 3 active neutrinos and $n$ sterile neutrinos in the
basis where the flavor eigenstates of the charged leptons are
identified with their mass eigenstates:
\begin{eqnarray}
\left(\begin{matrix} \nu^{}_e \cr \nu^{}_\mu \cr \nu^{}_\tau \cr \vdots \cr
\end{matrix}
\right) = \left(\begin{matrix} V^{}_{e1} & V^{}_{e2} & V^{}_{e3} & \cdots \cr
V^{}_{\mu 1} & V^{}_{\mu 2} & V^{}_{\mu 3} & \cdots \cr
V^{}_{\tau 1} & V^{}_{\tau 2} & V^{}_{\tau 3} & \cdots \cr
\vdots & \vdots & \vdots & \ddots \cr \end{matrix} \right)
\left(\begin{matrix} \nu^{}_1 \cr \nu^{}_2 \cr \nu^{}_3 \cr \vdots \cr
\end{matrix}
\right) \; .
\end{eqnarray}
Hence the $3\times 3$ MNSP matrix itself must be non-unitary. From
the point of view of neutrino oscillations, one may classify its
possible non-unitary effects into three categories:
\begin{itemize}
\item     the {\it indirect} non-unitary effect arising from the heavy
sterile neutrinos which are kinematically forbidden to take part in
neutrino oscillations;

\item     the {\it direct} non-unitary effect caused by the light sterile
neutrinos which are able to participate in neutrino oscillations;

\item     the {\it interplay} of the direct and indirect non-unitary
effects in a flavor mixing scenario including both light and heavy
sterile neutrinos.
\end{itemize}
In each of the three cases, no matter how small or how large the
mass scale of sterile neutrinos could be, the experimental
information on the matrix elements of $V$ is essentially different
from that in the standard case (i.e., the case in which $V$ is
exactly a $3\times 3$ unitary matrix). Hence testing the unitarity
of the $3\times 3$ MNSP matrix is experimentally important to
constrain the flavor mixing parameters of possible new physics and
can theoretically shed light on the underlying dynamics responsible
for the neutrino mass generation and lepton flavor mixing (e.g., the
$3\times 3$ MNSP matrix is exactly unitary in the type-II \cite{SS2}
seesaw mechanism but non-unitary in the type-I \cite{SS} and
type-III \cite{SS3} seesaw mechanisms).

Following a similar strategy for the precision test of the unitarity
of the $3\times 3$ Cabibbo-Kobayashi-Maskawa (CKM) quark flavor
mixing matrix \cite{PDG}, here we concentrate on a possible
experimental test of the normalization relation $|V^{}_{e1}|^2 +
|V^{}_{e2}|^2 + |V^{}_{e3}|^2 =1$ of the $3\times 3$ MNSP matrix in
a precision reactor experiment which is expected to be able to
distinguish between the oscillation modes induced by $\Delta
m^2_{31}$ and $\Delta m^2_{32}$. The recently proposed Daya Bay II
reactor antineutrino oscillation experiment \cite{DYB2} is just of
this type, so is the proposed RENO-50 reactor experiment
\cite{RENO50}. At present a very preliminary constraint on the sum
of $|V^{}_{e1}|^2$, $|V^{}_{e2}|^2$ and $|V^{}_{e3}|^2$ is
\cite{Antusch}
\begin{eqnarray}
|V^{}_{e1}|^2 + |V^{}_{e2}|^2 + |V^{}_{e3}|^2 = 0.994 \pm 0.005 \;
\end{eqnarray}
at the $90\%$ confidence level, implying that the $3\times 3$ MNSP
matrix is allowed to be non-unitary at the $\lesssim 1\%$ level. In
this note we shall discuss how to determine $|V^{}_{e1}|^2$,
$|V^{}_{e2}|^2$ and $|V^{}_{e3}|^2$ via a precision measurement of
the $\overline{\nu}^{}_e \to \overline{\nu}^{}_e$ oscillation and
examine whether their sum deviates from one or not. To be more
specific, we are going to consider three typical scenarios of
active-sterile neutrino mixing to illustrate possible non-unitary
effects on the $3\times 3$ MNSP matrix as listed above:
\begin{itemize}
\item     The (3+{\bf 3}) flavor mixing scenario with three heavy sterile
neutrinos which indirectly violate the unitarity of the $3\times 3$
MNSP matrix;

\item     The (3+$\mathbbm{1}$) flavor mixing scenario with a single light
sterile neutrino which directly violates the unitarity of the $3\times 3$
MNSP matrix;

\item     The (3+$\mathbbm{1}$+{\bf 2}) flavor mixing scenario in which
the light and heavy sterile neutrinos violate the unitarity of the
$3\times 3$ MNSP matrix directly and indirectly, respectively.
\end{itemize}
In each case the sum $|V^{}_{e1}|^2 + |V^{}_{e2}|^2 + |V^{}_{e3}|^2$
can be expressed in terms of the active-sterile neutrino mixing
angles in a given parametrization of the overall $(3+n) \times
(3+n)$ flavor mixing matrix. Taking the parametrization proposed in
Ref. \cite{Xing12} for example, we shall show that it is possible to
determine the active neutrino mixing angles $\theta^{}_{12}$ and
$\theta^{}_{13}$ without any contamination coming from the sterile
neutrinos. We hope that the points to be addressed in the remaining
part of this note will be helpful for the next-generation reactor
antineutrino oscillation experiments, either to test the standard
$3\times 3$ MNSP flavor mixing picture or to probe new physics via
its possible non-unitary effects.

\section{Non-unitary effects}

In the presence of $n$ species of sterile neutrinos, no matter
whether they are very light or very heavy, the amplitude of the
active $\nu^{}_\alpha \to \nu^{}_\beta$ oscillation (for $\alpha,
\beta = e, \mu, \tau$) in vacuum can be expressed as
\begin{eqnarray}
A(\nu^{}_\alpha \to \nu^{}_\beta) & \hspace{-0.2cm} =
\hspace{-0.2cm} & \sum_i \left[A(W^+ \to l^+_\alpha \nu^{}_i) \cdot
{\rm Prop}(\nu^{}_i) \cdot A(\nu^{}_i W^- \to l^-_\beta) \right]
\nonumber \\
& \hspace{-0.2cm} = \hspace{-0.2cm} &
\frac{1}{\sqrt{(VV^\dagger)^{}_{\alpha\alpha}
(VV^\dagger)^{}_{\beta\beta}}} \sum_i \left[V^*_{\alpha i}
\exp\left(\displaystyle -i\frac{m^2_i L}{2E}\right) V^{}_{\beta
i}\right] \; ,
\end{eqnarray}
in which $A(W^+\to l^+_\alpha \nu^{}_i) = V^*_{\alpha
i}/\sqrt{(VV^\dagger)_{\alpha\alpha}}~$, ${\rm Prop}(\nu^{}_i)$ and
$A(\nu^{}_i W^-\to l^-_\beta) = V^{}_{\beta
i}/\sqrt{(VV^\dagger)_{\beta\beta}}$ describe the production of
$\nu^{}_\alpha$ via the weak charged-current interaction, the
propagation of free $\nu^{}_i$ and the detection of $\nu^{}_\beta$
via the weak charged-current interaction, respectively
\cite{Antusch}---\cite{Giunti} (a schematic diagram is shown in Fig.
1 for illustration). Here $m^{}_i$ is the mass of the light (active
or sterile) neutrino $\nu^{}_i$, $E$ denotes the neutrino beam
energy and $L$ stands for the distance between the source and the
detector. It is then straightforward to calculate the probability
$P(\nu^{}_\alpha \to \nu^{}_\beta) = |A(\nu^{}_\alpha \to
\nu^{}_\beta)|^2$. We obtain
\footnote{Note that $(VV^\dagger)^{}_{\alpha\alpha}$ and
$(VV^\dagger)^{}_{\beta\beta}$ in front of $P(\nu^{}_\alpha \to
\nu^{}_\beta)$ are the normalization factors and can essentially be
canceled by the same factors coming from the production of
$\nu^{}_\alpha$ at the source and the detection of $\nu^{}_\beta$ at
the detector \cite{Antusch}, which are both governed by the weak
charged-current interactions. Hence we focus on ${\cal
P}(\nu^{}_\alpha \to \nu^{}_\beta)$ in this work.}
\begin{eqnarray}
{\cal P}(\nu^{}_\alpha \to \nu^{}_\beta) & \hspace{-0.2cm} \equiv
\hspace{-0.2cm} & (VV^\dagger)^{}_{\alpha\alpha} \cdot
P(\nu^{}_\alpha \to \nu^{}_\beta) \cdot (VV^\dagger)^{}_{\beta\beta}
\nonumber \\
& \hspace{-0.2cm} = \hspace{-0.2cm} & \sum^{}_i |V^{}_{\alpha i}|^2
|V^{}_{\beta i}|^2 + 2 \sum^{}_{i<j} \left[ {\rm Re} \left(
V^{}_{\alpha i} V^{}_{\beta j} V^*_{\alpha j} V^*_{\beta i} \right)
\cos \Delta^{}_{ij} - {\rm Im} \left( V^{}_{\alpha i} V^{}_{\beta j}
V^*_{\alpha j} V^*_{\beta i} \right) \sin\Delta^{}_{ij} \right] \; ,
~~~~~~~~
\end{eqnarray}
where $\Delta^{}_{ij} \equiv \Delta m^2_{ij} L/(2E)$ with $\Delta
m^2_{ij} \equiv m^2_i - m^2_j$. One may write out the expression of
${\cal P}(\overline{\nu}^{}_\alpha \to \overline{\nu}^{}_\beta)$ or
$P(\overline{\nu}^{}_\alpha \to \overline{\nu}^{}_\beta)$ from Eq.
(4) by making the replacement $V \Longrightarrow V^*$. If $V$ is
exactly unitary, then $(VV^\dagger)^{}_{\alpha\alpha} =
(VV^\dagger)^{}_{\beta\beta} = 1$ holds and thus ${\cal
P}(\nu^{}_\alpha \to \nu^{}_\beta) = P(\nu^{}_\alpha \to
\nu^{}_\beta)$ is just the conventional formula of $\nu^{}_\alpha
\to \nu^{}_\beta$ oscillations. Here we only pay interest to the
reactor antineutrino oscillations in vacuum,
\begin{eqnarray}
{\cal P}(\overline{\nu}^{}_e \to \overline{\nu}^{}_e) =
\left(\sum_i |V^{}_{e i}|^2 \right)^2 - 4 \sum^{}_{i<j}
\left(|V^{}_{e i}|^2 |V^{}_{e j}|^2 \sin^2
\frac{\Delta m^{2}_{ij} L}{4 E} \right) \; .
\end{eqnarray}
Note that the energy-independent term on the right-hand side of Eq.
(5) is not equal to one if there are heavy sterile antineutrinos
which do not take part in the oscillation. This point will be made
clear in the subsequent discussions, in which a few typical
active-sterile neutrino mixing scenarios will be taken into account.
Given the fact that the typical value of $E$ is a few MeV and that
of $L$ is usually less than 100 km for a realistic experiment, it is
completely unnecessary for us to consider the negligibly small
terrestrial matter effects on ${\cal P}(\overline{\nu}^{}_e \to
\overline{\nu}^{}_e)$.

\subsection{The $(3+{\bf 3})$ flavor mixing scenario}

Let us first consider the (3+{\bf 3}) flavor mixing scenario with
three heavy sterile neutrinos which indirectly violate the unitarity
of the $3\times 3$ MNSP matrix. In this case a full parametrization
of the $6\times 6$ flavor mixing matrix has been given in Ref.
\cite{Xing12}, and the elements in its first row read
\begin{eqnarray}
V^{}_{e1} & \hspace{-0.2cm} = \hspace{-0.2cm} & c^{}_{12} c^{}_{13}
c^{}_{14} c^{}_{15} c^{}_{16} \; ,
\nonumber \\
V^{}_{e2} & \hspace{-0.2cm} = \hspace{-0.2cm} & \hat{s}^*_{12}
c^{}_{13} c^{}_{14} c^{}_{15} c^{}_{16} \; ,
\nonumber \\
V^{}_{e3} & \hspace{-0.2cm} = \hspace{-0.2cm} & \hat{s}^*_{13}
c^{}_{14} c^{}_{15} c^{}_{16} \; ,
\nonumber \\
V^{}_{e 4} & \hspace{-0.2cm} = \hspace{-0.2cm} & \hat{s}^*_{14}
c^{}_{15} c^{}_{16} \; , \nonumber \\
V^{}_{e 5} & \hspace{-0.2cm} = \hspace{-0.2cm} & \hat{s}^*_{15}
c^{}_{16} \; , \nonumber \\
V^{}_{e 6} & \hspace{-0.2cm} = \hspace{-0.2cm} & \hat{s}^*_{16} \; ,
\end{eqnarray}
where $c^{}_{ij} \equiv \cos\theta^{}_{ij}$ and $\hat{s}^{}_{ij}
\equiv \sin\theta^{}_{ij} e^{i\delta^{}_{ij}}$ with $\theta^{}_{ij}$
being the flavor mixing angles and $\delta^{}_{ij}$ being the
CP-violating phases. In view of the fact that the active-sterile
neutrino mixing angles $\theta^{}_{14}$, $\theta^{}_{15}$ and
$\theta^{}_{16}$ are at most of ${\cal O}(0.1)$ in magnitude
\cite{Antusch}, the actual values of the three active neutrino
mixing angles $\theta^{}_{12}$, $\theta^{}_{13}$ and
$\theta^{}_{23}$ should be very close to those extracted from
current neutrino oscillation data by assuming the $3\times 3$ MNSP
matrix to be exactly unitary. Eq. (6) leads us to
\begin{eqnarray}
|V^{}_{e1}|^2 + |V^{}_{e2}|^2 + |V^{}_{e3}|^2 = c^2_{14} c^2_{15}
c^2_{16} \simeq 1 - \left(s^2_{14} + s^2_{15} + s^2_{16} \right) \;
,
\end{eqnarray}
implying that the sum of $|V^{}_{e1}|^2$, $|V^{}_{e2}|^2$ and
$|V^{}_{e3}|^2$ is possible to deviate from one either at the
percent level or at a much lower level. Is such a small effect
really detectable in a precision reactor $\overline{\nu}^{}_e \to
\overline{\nu}^{}_e$ oscillation experiment?

In the present scenario the three hypothetical heavy sterile
neutrinos are kinematically forbidden to participate in neutrino
oscillations. So Eq. (5) can be simplified to
\begin{eqnarray}
{\cal P}(\overline{\nu}^{}_e \to \overline{\nu}^{}_e)
& \hspace{-0.2cm} = \hspace{-0.2cm} & I^{}_0
- 4A \sin^2 \frac{\Delta m^{2}_{21} L}{4 E}
- 4B \sin^2 \frac{\Delta m^{2}_{31} L}{4 E} \nonumber \\
& & + 2C \sin \frac{\Delta m^{2}_{21} L}{2 E}
\sin \frac{\Delta m^{2}_{31} L}{2 E} +
8C \sin^2 \frac{\Delta m^{2}_{21} L}{4 E}
\sin^2 \frac{\Delta m^{2}_{31} L}{4 E} \; ,
\end{eqnarray}
where $\Delta m^2_{32} = \Delta m^2_{31} - \Delta m^2_{21}$ has been
used, and
\begin{eqnarray}
I^{}_0 & \hspace{-0.2cm} = \hspace{-0.2cm} &
\left(|V^{}_{e1}|^2 + |V^{}_{e2}|^2 + |V^{}_{e3}|^2\right)^2
\; , \nonumber \\
A & \hspace{-0.2cm} = \hspace{-0.2cm} &
\left( |V^{}_{e1}|^2 + |V^{}_{e3}|^2 \right) |V^{}_{e2}|^2
\; , \nonumber \\
B & \hspace{-0.2cm} = \hspace{-0.2cm} &
\left( |V^{}_{e1}|^2 + |V^{}_{e2}|^2 \right) |V^{}_{e3}|^2
\; , \nonumber \\
C & \hspace{-0.2cm} = \hspace{-0.2cm} &
|V^{}_{e2}|^2 |V^{}_{e3}|^2 \; .
\end{eqnarray}
Note that the third oscillation term in Eq. (8) is sensitive to the
unknown sign of $\Delta m^2_{31}$ (i.e., $\Delta m^2_{31} >0$ and
$\Delta m^2_{31} <0$ correspond to the normal and inverted neutrino
mass hierarchies, respectively), and it might be measurable in a
precision reactor antineutrino oscillation experiment in the
foreseeable future. Some comments on the implications of Eqs. (8)
and (9) are in order.
\begin{itemize}
\item     The non-unitarity of the $3\times 3$ MNSP matrix, characterized
by the small deviation of $I^{}_0$ from one, can be tested via the
energy-independent zero-distance effect \cite{LL} in such a
disappearance antineutrino oscillation experiment:
\begin{eqnarray}
{\cal P}(\overline{\nu}^{}_e \to \overline{\nu}^{}_e)\left|^{}_{L =0}
\right . = I^{}_0
= c^4_{14} c^4_{15} c^4_{16} \simeq 1 - 2 \left(s^2_{14} + s^2_{15}
+ s^2_{16} \right) \; .
\end{eqnarray}
This effect is in principle measurable with the help of a near
detector, although it might in practice be indistinguishable from
the background which at least includes the uncertainties associated
with the antineutrino flux. If the energy spectrum of the
$\overline{\nu}^{}_e \to \overline{\nu}^{}_e$ oscillation can be
fully established, however, it should be possible to determine or
constrain the size of $I^{}_0$.

\item     The flavor mixing factors $A$, $B$ and $C$ are
simple functions of the matrix elements $|V^{}_{e1}|^2$,
$|V^{}_{e2}|^2$ and $|V^{}_{e3}|^2$. It is therefore straightforward
to obtain
\begin{eqnarray}
|V^{}_{e1}|^2 & \hspace{-0.2cm} = \hspace{-0.2cm} &
\sqrt{\frac{\left(A - C\right) \left(B - C\right)}{C}} \; ,
\nonumber \\
|V^{}_{e2}|^2 & \hspace{-0.2cm} = \hspace{-0.2cm} &
\sqrt{\frac{\left(A - C\right) C}{B - C}} \; ,
\nonumber \\
|V^{}_{e3}|^2 & \hspace{-0.2cm} = \hspace{-0.2cm} &
\sqrt{\frac{\left(B - C\right) C}{A - C}} \; .
\end{eqnarray}
Provided $A$, $B$ and $C$ are all measured to a good degree of
accuracy, one may also use Eq. (11) to calculate the sum
$|V^{}_{e1}|^2 + |V^{}_{e2}|^2 + |V^{}_{e3}|^2 = \sqrt{I^{}_0}$ and
examine whether it departs from one or not. On the other hand, the
value of $I^{}_0$ to be determined in this way can be compared with
the one to be measured from the zero-distance effect as shown in Eq.
(10).

\item     Given the concise parametrization in Eq. (6), the active
neutrino mixing angles $\theta^{}_{12}$ and $\theta^{}_{13}$ can be
determined from the ratios $A/I^{}_0$, $B/I^{}_0$ and $C/I^{}_0$
without any contamination coming from the heavy sterile neutrinos.
This point is clearly seen as follows:
\begin{eqnarray}
\frac{A}{I^{}_0} & \hspace{-0.2cm} = \hspace{-0.2cm} & \frac{1}{4}
\left( \sin^2 2\theta^{}_{12} \cos^4 \theta^{}_{13}
+ \sin^2 \theta^{}_{12} \sin^2 2\theta^{}_{13} \right) \; ,
\nonumber \\
\frac{B}{I^{}_0} & \hspace{-0.2cm} = \hspace{-0.2cm} &
\frac{1}{4} \sin^2 2\theta^{}_{13} \; ,
\nonumber \\
\frac{C}{I^{}_0} & \hspace{-0.2cm} = \hspace{-0.2cm} & \frac{1}{4}
\sin^2 \theta^{}_{12} \sin^2 2\theta^{}_{13} \; .
\end{eqnarray}
Taking $\theta^{}_{12} \simeq 34^\circ$ and $\theta^{}_{13} \simeq
9^\circ$ for example \cite{FIT}, we immediately have $A/I^{}_0
\simeq 0.212$, $B/I^{}_0 \simeq 0.024$ and $C/I^{}_0 \simeq 0.0075$.
The smallness of $C$ makes it rather difficult to be measured (in
other words, a determination of the sign of $\Delta m^2_{31}$ or the
neutrino mass ordering is a big challenge to the reactor
antineutrino oscillation experiments \cite{Petcov}---\cite{Zhang}).
\end{itemize}
To observe the non-unitary effects in the (3+{\bf 3}) flavor mixing
scenario under consideration, one may also study the appearance
neutrino oscillation (e.g., the $\nu^{}_\mu \to \nu^{}_\tau$
oscillation) experiments with reasonably long baselines. In such a
precision accelerator neutrino oscillation experiment, even new
CP-violating effects are likely to show up at the ${\cal
O}(10^{-3})$ or ${\cal O}(10^{-2})$ level provided the relevant
active-sterile flavor mixing angles are not strongly suppressed
\cite{Yasuda}.

At this point it is worth mentioning that the type-I seesaw mechanism
around the TeV energy scale can simply lead to the (3+{\bf 3}) flavor
mixing scenario with appreciable non-unitary effects \cite{X09}.
Similar effects may result from some interesting extensions of the
type-I seesaw mechanism, such as the inverse seesaw scenario \cite{MV},
the linear seesaw scenario \cite{LS} and the multiple seesaw
scenarios \cite{XZ}. In such model-building exercises, presumably
significant lepton-flavor-violating effects and non-standard neutrino
interactions are also expected to show up and their interplay with
the non-unitary effects offers an important and complementary window
for probing the true mechanism of neutrino mass generation and lepton
flavor mixing \cite{Valle}. On the experimental side, the precision
reactor antineutrino oscillation experiments will be complementary to
the precision accelerator neutrino experiments for our physics goal
and thus deserve particular attention.

\subsection{The $(3+\mathbbm{1})$ flavor mixing scenario}

To illustrate the direct non-unitary effect on the $3\times 3$ MNSP
matrix, we consider the (3+$\mathbbm{1}$) flavor mixing scenario
with a single light sterile neutrino motivated by the LSND
\cite{LSND}, MiniBooNE \cite{M} and reactor \cite{R} antineutrino
anomalies. In this case we have $|V^{}_{e1}|^2 + |V^{}_{e2}|^2 +
|V^{}_{e3}|^2 = 1 - |V^{}_{e4}|^2 = \cos^2 \theta^{}_{14}$ by using
the same parametrization as used above \cite{Xing12}, where
$\theta^{}_{14}$ is the active-sterile neutrino mixing angle.
Because the light sterile antineutrino takes part in the
$\overline{\nu}^{}_e \to \overline{\nu}^{}_e$ oscillation, Eq. (5)
is now simplified to
\begin{eqnarray}
{\cal P}(\overline{\nu}^{}_e \to \overline{\nu}^{}_e)
& \hspace{-0.2cm} = \hspace{-0.2cm} &
1 - 4 A^\prime \sin^2 \frac{\Delta m^{2}_{21} L}{4 E}
- 4 B^\prime \sin^2 \frac{\Delta m^{2}_{31} L}{4 E}
- 4 X^\prime \sin^2 \frac{\Delta m^{2}_{41} L}{4 E}
\nonumber \\
& & + 2 C^\prime \sin \frac{\Delta m^{2}_{21} L}{2 E}
\sin \frac{\Delta m^{2}_{31} L}{2 E} +
8 C^\prime \sin^2 \frac{\Delta m^{2}_{21} L}{4 E}
\sin^2 \frac{\Delta m^{2}_{31} L}{4 E}
\nonumber \\
& & + 2 Y^\prime \sin \frac{\Delta m^{2}_{21} L}{2 E}
\sin \frac{\Delta m^{2}_{41} L}{2 E} +
8 Y^\prime \sin^2 \frac{\Delta m^{2}_{21} L}{4 E}
\sin^2 \frac{\Delta m^{2}_{41} L}{4 E}
\nonumber \\
& & + 2 Z^\prime \sin \frac{\Delta m^{2}_{31} L}{2 E}
\sin \frac{\Delta m^{2}_{41} L}{2 E} +
8 Z^\prime \sin^2 \frac{\Delta m^{2}_{31} L}{4 E}
\sin^2 \frac{\Delta m^{2}_{41} L}{4 E} \; ,
\end{eqnarray}
where
\begin{eqnarray}
A^\prime & \hspace{-0.2cm} = \hspace{-0.2cm} &
\left( 1 - |V^{}_{e2}|^2 \right) |V^{}_{e2}|^2
= \frac{1}{4} \left( \sin^2 2\theta^{}_{12} \cos^4 \theta^{}_{13}
+ \sin^2 \theta^{}_{12} \sin^2 2\theta^{}_{13} \right)
\cos^4\theta^{}_{14} + \frac{1}{4} \sin^2\theta^{}_{12}
\cos^2\theta^{}_{13} \sin^2 2\theta^{}_{14} \; ,
\nonumber \\
B^\prime & \hspace{-0.2cm} = \hspace{-0.2cm} & \left( 1 -
|V^{}_{e3}|^2 \right) |V^{}_{e3}|^2 = \frac{1}{4} \sin^2 2\theta^{}_{13} \cos^4\theta^{}_{14} + \frac{1}{4} \sin^2\theta^{}_{13}
\sin^2 2\theta^{}_{14} \; ,
\nonumber \\
C^\prime & \hspace{-0.2cm} = \hspace{-0.2cm} & |V^{}_{e2}|^2
|V^{}_{e3}|^2 = \frac{1}{4}
\sin^2 \theta^{}_{12} \sin^2 2\theta^{}_{13} \cos^4\theta^{}_{14} \; ,
\nonumber \\
X^\prime & \hspace{-0.2cm} = \hspace{-0.2cm} & \left( 1 -
|V^{}_{e4}|^2 \right) |V^{}_{e4}|^2 = \frac{1}{4} \sin^2
2\theta^{}_{14} \; ,
\nonumber \\
Y^\prime & \hspace{-0.2cm} = \hspace{-0.2cm} &
|V^{}_{e2}|^2 |V^{}_{e4}|^2 = \frac{1}{4} \sin^2\theta^{}_{12}
\cos^2\theta^{}_{13} \sin^2 2\theta^{}_{14} \; ,
\nonumber \\
Z^\prime & \hspace{-0.2cm} = \hspace{-0.2cm} & |V^{}_{e3}|^2
|V^{}_{e4}|^2 = \frac{1}{4} \sin^2 \theta^{}_{13} \sin^2
2\theta^{}_{14} \; .
\end{eqnarray}
It is obvious that the oscillation terms associated with $2C^\prime$
and $2Y^\prime$ (or $2Z^\prime$) are sensitive to the signs of
$\Delta m^2_{31}$ and $\Delta m^2_{41}$ (or both of them),
respectively. This observation makes sense for the place of the
sterile neutrino in the whole mass spectrum because its mass
$m^{}_4$ is in general unnecessary to be larger than the mass scale of the
three active neutrinos (in particular if they are nearly degenerate
and close to 1 eV). A simple exercise leads us to the solutions
\footnote{Note that the other solution of $|V^{}_{ei}|^2$ (for
$i=2,3,4$) is expected to be inconsistent with the observed neutrino
mixing pattern, and thus it is ignored here.}
\begin{eqnarray}
|V^{}_{e2}|^2 & \hspace{-0.2cm} = \hspace{-0.2cm} &
\frac{1}{2} \left(1 - \sqrt{1 - 4 A^\prime} \right) \; ,
\nonumber \\
|V^{}_{e3}|^2 & \hspace{-0.2cm} = \hspace{-0.2cm} &
\frac{1}{2} \left(1 - \sqrt{1 - 4 B^\prime} \right) \; ,
\nonumber \\
|V^{}_{e4}|^2 & \hspace{-0.2cm} = \hspace{-0.2cm} &
\frac{1}{2} \left(1 - \sqrt{1 - 4 X^\prime} \right) \; .
\end{eqnarray}
One may also obtain
$|V^{}_{e4}|^2 = \sqrt{Y^\prime Z^\prime/C^\prime}$ and
some interesting correlations such as
\begin{eqnarray}
C^\prime & \hspace{-0.2cm} = \hspace{-0.2cm} &
\frac{1}{4} \left(1 - \sqrt{1 - 4A^\prime} \right)
\left(1 - \sqrt{1 - 4B^\prime} \right) \; ,
\nonumber \\
\frac{Y^\prime}{Z^\prime} & \hspace{-0.2cm} = \hspace{-0.2cm} &
\frac{1 - \sqrt{1 - 4A^\prime}}
{1 - \sqrt{1 - 4B^\prime}} \; .
\end{eqnarray}
A few comments on the above results are in order.
\begin{itemize}
\item     Given a light sterile antineutrino whose absolute mass
scale is unspecified or $\Delta m^2_{41}$ is unknown, it is in
principle possible to determine or constrain the active-sterile
flavor mixing angle $\theta^{}_{14}$ via a precision measurement of
$A^\prime$, $B^\prime$ and $C^\prime$. The deviation of these three
parameters from their values in the standard case is characterized by
nonzero $\sin^2\theta^{}_{14}$.
Taking $\theta^{}_{12} \simeq 34^\circ$, $\theta^{}_{13} \simeq
9^\circ$ and $\theta^{}_{14} \simeq 10^\circ$ as a typical example,
we obtain $A^\prime \simeq 0.208$, $B^\prime \simeq 0.023$ and
$C^\prime \simeq 0.0071$, which can be compared with $A^\prime \simeq
0.212$, $B^\prime \simeq 0.024$ and $C^\prime \simeq 0.0075$ in
the $\theta^{}_{14} \simeq 0^\circ$ case.

\item     Of course, a direct measurement of the oscillation term
driven by $\Delta m^2_{41}$ will provide the direct evidence for the
existence of a light sterile antineutrino in the
$\overline{\nu}^{}_e \to \overline{\nu}^{}_e$ oscillation. The most
optimistic case, which seems to be very unlikely, is that
$X^\prime$, $Y^\prime$ and $Z^\prime$ could all be determined.
Considering $\theta^{}_{12} \simeq 34^\circ$, $\theta^{}_{13} \simeq
9^\circ$ and $\theta^{}_{14} \simeq 10^\circ$, we have $X^\prime
\simeq 0.029$, $Y^\prime \simeq 0.0089$ and $Z^\prime \simeq
0.00072$. One can see that $Z^\prime$ is too small to be measured,
but it does not matter because the sign of $\Delta m^2_{41}$ is
essentially determinable from a precision measurement of the
interference term associated with $2Y^\prime$ in Eq. (13). In
particular, it should be noted that $Y^\prime/Z^\prime =
\sin^2\theta^{}_{12} \cot^2\theta^{}_{13} \simeq 12$ is a result
independent of the input value of $\theta^{}_{14}$.

\item     Provided $\Delta m^2_{4i} \gg \Delta m^2_{31} \simeq 2.5 \times
10^{-3} {\rm eV}^2$ holds (for $i=1,2,3$) and the experimental
baseline length $L$ favors the oscillation terms driven by $\Delta
m^2_{31}$ and $\Delta m^2_{32}$, the corresponding active-sterile
antineutrino oscillation terms $\sin^2 [\Delta m^2_{4i} L/(4E)]$
will practically be averaged to $1/2$ because they oscillate too
quickly. In this case the overall effect induced by the sterile
antineutrino becomes the zero-distance effect:
\begin{eqnarray}
{\cal P}(\overline{\nu}^{}_e \to \overline{\nu}^{}_e)\left|^{}_{L
=0} \right . = 1 - 2 \left(1 - |V^{}_{e4}|^2 \right) |V^{}_{e4}|^2 =
1- \frac{1}{2} \sin^2 2\theta^{}_{14} \; .
\end{eqnarray}
A near detector is certainly difficult to measure such a small
effect because of the uncertainties associated with the antineutrino
flux. Nevertheless, we remark that the similar effect could be
cross-checked from a precision determination of the flavor mixing
factors $A^\prime$, $B^\prime$ and $C^\prime$.
\end{itemize}
In the same way one may discuss an analogous flavor mixing scenario
with two or three light sterile antineutrinos which take part in the
$\overline{\nu}^{}_e \to \overline{\nu}^{}_e$ oscillation.

\subsection{The $(3+\mathbbm{1}+{\bf 2})$ flavor mixing scenario}

Given three sterile neutrinos, one of them can be assumed to be
light enough so as to either interpret the existing sub-eV
antineutrino anomalies or account for the keV warm dark matter
\cite{Sterile}. In such a (3+$\mathbbm{1}$+{\bf 2}) flavor mixing
scenario the two heavy sterile neutrinos may play the role in
realizing the seesaw and leptogenesis mechanisms \cite{Glashow}. The
deviation of the sum $|V^{}_{e1}|^2 + |V^{}_{e2}|^2 + |V^{}_{e3}|^2$
from one is also described by Eq. (6), but the $\overline{\nu}^{}_e
\to \overline{\nu}^{}_e$ oscillation probability turns out to be
\begin{eqnarray}
{\cal P}(\overline{\nu}^{}_e \to \overline{\nu}^{}_e)
& \hspace{-0.2cm} = \hspace{-0.2cm} &
\widehat{I}^{}_0
- 4 \widehat{A} \sin^2 \frac{\Delta m^{2}_{21} L}{4 E}
- 4 \widehat{B} \sin^2 \frac{\Delta m^{2}_{31} L}{4 E}
- 4 \widehat{X} \sin^2 \frac{\Delta m^{2}_{41} L}{4 E}
\nonumber \\
& & + 2 \widehat{C} \sin \frac{\Delta m^{2}_{21} L}{2 E}
\sin \frac{\Delta m^{2}_{31} L}{2 E} +
8 \widehat{C} \sin^2 \frac{\Delta m^{2}_{21} L}{4 E}
\sin^2 \frac{\Delta m^{2}_{31} L}{4 E}
\nonumber \\
& & + 2 \widehat{Y} \sin \frac{\Delta m^{2}_{21} L}{2 E}
\sin \frac{\Delta m^{2}_{41} L}{2 E} +
8 \widehat{Y} \sin^2 \frac{\Delta m^{2}_{21} L}{4 E}
\sin^2 \frac{\Delta m^{2}_{41} L}{4 E}
\nonumber \\
& & + 2 \widehat{Z} \sin \frac{\Delta m^{2}_{31} L}{2 E}
\sin \frac{\Delta m^{2}_{41} L}{2 E} +
8 \widehat{Z} \sin^2 \frac{\Delta m^{2}_{31} L}{4 E}
\sin^2 \frac{\Delta m^{2}_{41} L}{4 E} \; ,
\end{eqnarray}
in which
\begin{eqnarray}
\widehat{I}^{}_0 & \hspace{-0.2cm} = \hspace{-0.2cm} &
\left(1 - |V^{}_{e5}|^2 - |V^{}_{e6}|^2\right)^2 \; ,
\nonumber \\
\widehat{A} & \hspace{-0.2cm} = \hspace{-0.2cm} &
\left(|V^{}_{e1}|^2 + |V^{}_{e3}|^2 + |V^{}_{e4}|^2 \right)
|V^{}_{e2}|^2 \; ,
\nonumber \\
\widehat{B} & \hspace{-0.2cm} = \hspace{-0.2cm} &
\left(|V^{}_{e1}|^2 + |V^{}_{e2}|^2 + |V^{}_{e4}|^2 \right)
|V^{}_{e3}|^2 \; ,
\nonumber \\
\widehat{C} & \hspace{-0.2cm} = \hspace{-0.2cm} &
|V^{}_{e2}|^2 |V^{}_{e3}|^2 \; ,
\nonumber \\
\widehat{X} & \hspace{-0.2cm} = \hspace{-0.2cm} &
\left(|V^{}_{e1}|^2 + |V^{}_{e2}|^2 + |V^{}_{e3}|^2\right)
|V^{}_{e4}|^2 \; ,
\nonumber \\
\widehat{Y} & \hspace{-0.2cm} = \hspace{-0.2cm} &
|V^{}_{e2}|^2 |V^{}_{e4}|^2 \; ,
\nonumber \\
\widehat{Z} & \hspace{-0.2cm} = \hspace{-0.2cm} &
|V^{}_{e3}|^2 |V^{}_{e4}|^2 \; .
\end{eqnarray}
We see that Eq. (18) consists of both direct and indirect
non-unitary effects on the $3\times 3$ MNSP matrix. A
straightforward calculation allows us to express $|V^{}_{ei}|^2$
(for $i=1,2,3,4$) in terms of the flavor mixing factors
$(\widehat{A}, \widehat{B}, \widehat{C})$ and $(\widehat{X},
\widehat{Y}, \widehat{Z})$ as follows: \small
\begin{eqnarray}
\normalsize |V^{}_{e1}|^2 & \hspace{-0.2cm} = \hspace{-0.2cm} &
\frac{\sqrt{\left(\widehat{A} - \widehat{C} - \widehat{Y}\right)
\left(\widehat{B} - \widehat{C} - \widehat{Z}\right) \widehat{C}}}
{\widehat{C}} \; ,
\nonumber \\
|V^{}_{e2}|^2 & \hspace{-0.2cm} = \hspace{-0.2cm} &
\frac{\sqrt{\left(\widehat{A} - \widehat{C} - \widehat{Y}\right)
\left(\widehat{B} - \widehat{C} - \widehat{Z}\right) \widehat{C}}}
{\widehat{B} - \widehat{C} - \widehat{Z}} \; ,
\nonumber \\
|V^{}_{e3}|^2 & \hspace{-0.2cm} = \hspace{-0.2cm} &
\frac{\sqrt{\left(\widehat{A} - \widehat{C} - \widehat{Y}\right)
\left(\widehat{B} - \widehat{C} - \widehat{Z}\right)
\widehat{C}}}{\widehat{A} - \widehat{C} - \widehat{Y}} \; ,
\nonumber \\
|V^{}_{e4}|^2 & \hspace{-0.2cm} = \hspace{-0.2cm} &
\frac{\widehat{X}
\sqrt{\left(\widehat{A} - \widehat{C} - \widehat{Y}\right)
\left(\widehat{B} - \widehat{C} - \widehat{Z}\right) \widehat{C}}}
{\left(\widehat{A} - \widehat{Y}\right)
\left(\widehat{B} - \widehat{Z}\right) - \widehat{C}^2} \; .
\end{eqnarray}
\normalsize Of course, $|V^{}_{e4}|^2 = \sqrt{\widehat{Y}
\widehat{Z}/\widehat{C}}$ holds too
\footnote{Note that the expressions obtained in Eq. (20) may also
cover the case of the (3+$\mathbbm{1}$) flavor mixing scenario
discussed in section 2.2. Namely, $|V^{}_{ei}|^2$ can be expressed
in terms of $(A^\prime, B^\prime, C^\prime)$ and $(X^\prime,
Y^\prime, Z^\prime)$ in the same form as Eq. (20).}.
Using the neutrino mixing angles given in Eq. (6), we obtain
$\widehat{I}^{}_0 = c^4_{15} c^4_{16}$ as well as
\begin{eqnarray}
\frac{\widehat{A}}{\widehat{I}^{}_0}
& \hspace{-0.2cm} = \hspace{-0.2cm} &
\frac{1}{4}
\left( \sin^2 2\theta^{}_{12} \cos^4 \theta^{}_{13}
+ \sin^2 \theta^{}_{12} \sin^2 2\theta^{}_{13} \right)
\cos^4\theta^{}_{14} + \frac{1}{4} \sin^2\theta^{}_{12}
\cos^2\theta^{}_{13} \sin^2 2\theta^{}_{14} \; ,
\nonumber \\
\frac{\widehat{B}}{\widehat{I}^{}_0}
& \hspace{-0.2cm} = \hspace{-0.2cm} &
\frac{1}{4} \sin^2 2\theta^{}_{13} \cos^4\theta^{}_{14} +
\frac{1}{4} \sin^2\theta^{}_{13} \sin^2 2\theta^{}_{14}
\; , \nonumber \\
\frac{\widehat{C}}{\widehat{I}^{}_0}
& \hspace{-0.2cm} = \hspace{-0.2cm} &
\frac{1}{4}
\sin^2 \theta^{}_{12} \sin^2 2\theta^{}_{13} \cos^4\theta^{}_{14} \; ,
\nonumber \\
\frac{\widehat{X}}{\widehat{I}^{}_0}
& \hspace{-0.2cm} = \hspace{-0.2cm} &
\frac{1}{4} \sin^2 2\theta^{}_{14} \; ,
\nonumber \\
\frac{\widehat{Y}}{\widehat{I}^{}_0}
& \hspace{-0.2cm} = \hspace{-0.2cm} &
\frac{1}{4} \sin^2\theta^{}_{12}
\cos^2\theta^{}_{13} \sin^2 2\theta^{}_{14} \; ,
\nonumber \\
\frac{\widehat{Z}}{\widehat{I}^{}_0}
& \hspace{-0.2cm} = \hspace{-0.2cm} &
\frac{1}{4} \sin^2 \theta^{}_{13} \sin^2
2\theta^{}_{14} \; ,
\end{eqnarray}
Some discussions and remarks are in order.
\begin{itemize}
\item     Switching off the light sterile neutrino (i.e.,
setting $|V^{}_{e4}|^2 = 0$ or equivalently $\widehat{X} =
\widehat{Y} = \widehat{Z} =0$), one can easily reproduce Eq. (11)
from Eq. (20). If the two heavy sterile neutrinos are switched off,
then it is straightforward to reproduce Eq. (14) from Eqs. (19) and
(21). The interplay between the direct non-unitary effect measured
by nonzero $\theta^{}_{14}$ and the indirect non-unitary effect
characterized by nonzero $\theta^{}_{15}$ and $\theta^{}_{16}$ is
therefore transparent. In particular, the relationship
\begin{eqnarray}
|V^{}_{e1}|^2 + |V^{}_{e2}|^2 + |V^{}_{e3}|^2 +
|V^{}_{e4}|^2 = \sqrt{\widehat{I}^{}_0} = c^2_{15} c^2_{16} \;
\end{eqnarray}
holds, and $\widehat{I}^{}_0$ can in principle be determined
from the zero-distance effect.

\item     Provided $\widehat{A}/\widehat{I}^{}_0$,
$\widehat{B}/\widehat{I}^{}_0$ and $\widehat{C}/\widehat{I}^{}_0$
are all measured in a precision reactor antineutrino experiment, it
should be possible to determine the flavor mixing angles
$\theta^{}_{12}$, $\theta^{}_{13}$ and $\theta^{}_{14}$ in a way
independent of the flavor mixing angles $\theta^{}_{15}$ and
$\theta^{}_{16}$. The same observation is true for the flavor mixing
factors $\widehat{X}/\widehat{I}^{}_0$,
$\widehat{Y}/\widehat{I}^{}_0$ and $\widehat{Z}/\widehat{I}^{}_0$,
although they must be much more difficult to be measured. Of course,
the situation might be simplified to some extent if the existence of
a light sterile antineutrino could be established somewhere else
(e.g., with the help of a suitable accelerator neutrino (or
antineutrino) oscillation experiment). But the important point is
that the reactor experiment can always provide some independent and
complementary information on the same physics, no matter whether it
is new or just conventional.

\item     In the present flavor mixing scenario, the effective neutrino
mass terms of the tritium beta decay $^3_1{\rm H} \to ~^3_2{\rm He}
+ e^- + \overline{\nu}^{}_e$ and the neutrinoless double-beta decay
$A(Z,A) \to A(Z+2, N-2) + 2 e^-$ can be expressed as
\begin{eqnarray}
\langle m\rangle^{\prime}_{e} & \hspace{-0.2cm} \equiv
\hspace{-0.2cm} & \left[ \sum_{i=1}^4 m^2_i |V^{}_{e i}|^2
\right]^{1/2} = \sqrt{\langle m\rangle^2_e c^2_{14} c^2_{15}
c^2_{16} + m^2_4 s^2_{14} c^2_{15} c^2_{16}} \;
\end{eqnarray}
with $\langle m\rangle^{}_e = \sqrt{m^2_1 c^2_{12} c^2_{13}
+ m^2_2 s^2_{12} c^2_{13} + m^2_3 s^2_{13}}$ being the standard
contribution from the three active neutrinos and
\begin{eqnarray}
\langle m\rangle^{\prime}_{ee} & \hspace{-0.2cm} \equiv
\hspace{-0.2cm} & \left| \sum_{i=1}^4 m^{}_i V^2_{e i} \right| =
\left| \langle
m\rangle^{}_{ee} \left(c^{}_{14} c^{}_{15} c^{}_{16} \right)^2 +
m^{}_4 \left(\hat{s}^*_{14} c^{}_{15} c^{}_{16}\right)^2 \right| \;
\end{eqnarray}
with $\langle m\rangle^{}_{ee} = m^{}_1 (c^{}_{12} c^{}_{13})^2 +
m^{}_2 (\hat{s}^*_{12} c^{}_{13})^2 + m^{}_3 (\hat{s}^*_{13})^2$
being the standard contribution from the active neutrinos,
respectively. Here we have assumed that a possible contribution
from the heavy sterile neutrinos is negligible \cite{Xing09}.
We see that $\langle m\rangle^\prime_e \geq \langle m\rangle^{}_e$
always holds, but it is difficult to judge the relative magnitudes of
$\left|\langle m\rangle^{}_{ee}\right|$ and
$\langle m\rangle^{\prime}_{ee}$ because the
relevant CP-violating phases may lead to more or less
(even complete) cancelations of different terms in them \cite{Xing03}.
\end{itemize}
If all the three sterile neutrinos are heavy enough, as in the
(3+{\bf 3}) flavor mixing scenario, we shall obtain the simpler
results $\langle m\rangle^\prime_e = \langle m\rangle^{}_e c{}_{14}
c^{}_{15} c^{}_{16}$ and $\langle m\rangle^\prime_{ee} = \left|\langle
m\rangle^{}_{ee} \right| c^2_{14} c^2_{15} c^2_{16}$.

\section{Concluding remarks}

In the era of precision neutrino physics one of the important jobs
is to test the unitarity of the $3\times 3$ MNSP flavor mixing
matrix and probe possible new physics which might give rise to some
observable non-unitary effects on it. Such effects serve as a special
example of possible non-standard interactions associated with massive
neutrinos and their oscillations \cite{TO}. Starting from this point
of view, we have classified three typical categories of non-unitary
effects on the MNSP matrix and illustrated them in a precision
reactor antineutrino oscillation experiment: 1) the {\it indirect}
effect in the (3+{\bf 3}) flavor mixing scenario where the three
heavy sterile neutrinos do not participate in neutrino oscillations;
2) the {\it direct} effect in the (3+$\mathbbm{1}$) scenario where
the light sterile neutrino can oscillate into the active ones; and
3) the {\it interplay} of both of them in the (3+$\mathbbm{1}$+{\bf
2}) scenario. We have shown that both the zero-distance effect and
flavor mixing factors of different oscillation modes can be used to
determine or constrain the sum of $|V^{}_{e1}|^2$, $|V^{}_{e2}|^2$
and $|V^{}_{e3}|^2$ and its possible deviation from one. In
addition, the active neutrino mixing angles $\theta^{}_{12}$ and
$\theta^{}_{13}$ can be cleanly extracted even in the presence of
light or heavy sterile neutrinos. Some useful analytical results,
which can be applied to a numerical analysis of the future experimental
data, have been presented for each of the three scenarios under
consideration.

We expect that some points of view in this work will be helpful for
a realistic reactor antineutrino oscillation experiment, such as the
proposed Daya Bay II experiment which aims to measure the neutrino
mass ordering and test the unitarity of the $3\times 3$ MNSP matrix.
We admit the big challenges that one has to face in practice, either
in measuring the zero-distance effect or in detecting different
oscillation modes, or in both of them. Nevertheless, we strongly
hope that the Daya Bay II experiment and other precision reactor
antineutrino oscillation experiments may play an important role in
probing possible new physics in the lepton sector, or can at least
be complementary to the precision accelerator neutrino oscillation
experiments in this respect in the foreseeable future.

An obvious drawback of this work is the lack of a numerical analysis
of the future experimental sensitivities to the potential non-unitary
effects on the MNSP matrix. This is simply because the main characteristics
of the currently proposed Daya Bay II and RENO-50 experiments remain
too preliminary and incomplete, and the existing constraints on the
non-unitary effects are also preliminary and more or less dependent
on the hypothetical active-sterile flavor mixing scheme.
In this case we have to focus on the analytical
discussions about the salient features of direct and indirect non-unitary
effects (and their interplay) in the present work, leaving a quantitative
study of the same topic in a future work.

\vspace{0.3cm}

The author is indebted to Y.F. Li and S. Luo for intensive and
valuable discussions, and in particular to S. Luo for her partial
involvement at the early stage of this work. He is also grateful to
C. Giunti and S. Zhou for helpful comments on the interplay of
direct and indirect non-unitary effects. This work is supported in
part by the National Natural Science Foundation of China under Grant
No. 11135009.


\newpage

\begin{figure}[t]
\vspace{-3cm}
\epsfig{file=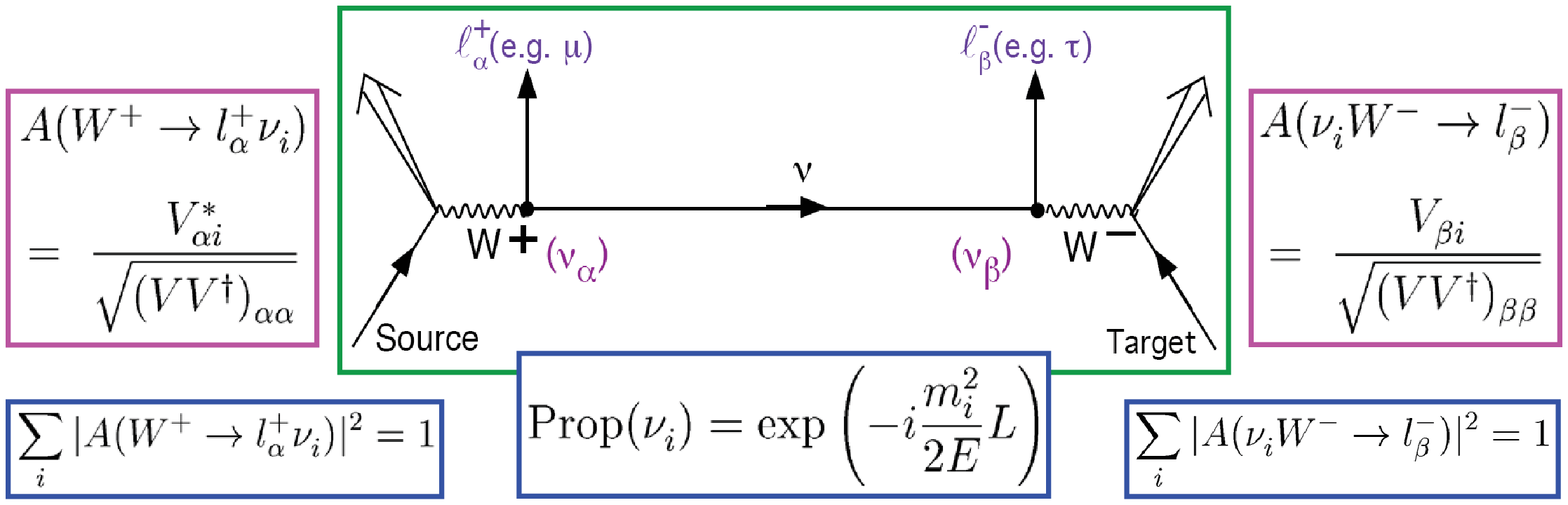,bbllx=1.3cm,bblly=14cm,bburx=10.5cm,bbury=20cm,%
width=7cm,angle=0,clip=0} \vspace{5.2cm} \caption{The schematic
diagram for the production of the $\nu^{}_\alpha$ neutrino via the
weak charged-current interaction, the propagation of free $\nu^{}_i$
neutrinos in vacuum and the detection of the $\nu^{}_\beta$ neutrino
via the weak charged-current interaction in the $\nu^{}_\alpha \to
\nu^{}_\beta$ oscillation process (for $\alpha,\beta =e, \mu,
\tau$), in which there might be possible non-unitary effects on the
$3\times 3$ MNSP flavor mixing matrix due to the presence of light
and (or) heavy sterile neutrinos.}
\end{figure}

\end{document}